\documentclass{article}
\usepackage{CJKutf8}
\usepackage{spconf,amsmath,graphicx}
\usepackage{paralist}
\usepackage{booktabs,multirow,makecell}
\usepackage{url}
\usepackage{cite}
\usepackage{hyperref}
\hypersetup{
    colorlinks=true,
    linkcolor=red,
    urlcolor=magenta,
    }

\title{Can Whisper perform speech-based in-context learning?}
\name{Siyin Wang$^1$, Chao-Han Yang$^2$, Ji Wu$^1$, Chao Zhang$^{1,\ast}$
\thanks{$\ast$ Corresponding author}
}
\address{
  $^1$Department of Electronic Engineering, Tsinghua University, Beijing, China \\
  $^2$School of Electrical and Computer Engineering, Georgia Institute of Technology, Atlanta, USA \\
\texttt{\small{wangsiyi23@mails.tsinghua.edu.cn; cz277@tsinghua.edu.cn}}
}
\begin{document}
\ninept
\maketitle
\begin{abstract}
This paper investigates the in-context learning abilities of the Whisper automatic speech recognition (ASR) models released by OpenAI. A novel speech-based in-context learning (SICL) approach is proposed for test-time adaptation, which can reduce the word error rates (WERs) with only a small number of labelled speech samples without gradient descent. Language-level adaptation experiments using Chinese dialects showed that when applying SICL to isolated word ASR, consistent and considerable relative WER reductions can be achieved using Whisper models of any size on two dialects, which is on average 32.3\%. A $k$-nearest-neighbours-based in-context example selection technique can be applied to further improve the efficiency of SICL, which can increase the average relative WER reduction to 36.4\%. The findings are verified using speaker adaptation or continuous speech recognition tasks, and both achieved considerable relative WER reductions. Detailed quantitative analyses are also provided to shed light on SICL's adaptability to phonological variances and dialect-specific lexical nuances.

\end{abstract}
\section{Introduction}

Recently, large language models (LLMs) \cite{C1,C28,C29} have propelled state-of-the-art performance across diverse natural language processing tasks. Through a prompting paradigm where only a handful of input-label pairings specify the task, LLMs can perform adaptation and inference, a phenomenon termed ``in-context learning'' (ICL) \cite{C1,C8, C31hu2022context}. While being widely studied in text and image models \cite{C2,C3,C4}, ICL has not yet been studied within the realm of speech processing. Given the inherent linguistic alignment between speech and text, some large-scale speech models \cite{C5,C6} might harness contextual cues for ICL, which can emerge as a pivotal research direction.

This paper presents the first study on ICL for speech processing to the best of our knowledge.   
OpenAI's Whisper models \cite{C5} are applied to Chinese dialect automatic speech recognition (ASR), which use the Transformer encoder-decoder structure \cite{C26} and can be regarded as LLMs grounded on speech inputs. 
%
We propose a speech-based ICL (SICL) approach that leverages several \textit{in-context examples} (paired spoken words and labels) from a specific dialect or speaker to achieve test-time language- or speaker-level model adaptation without gradient descent. 
Compared to ICL for text-based LLMs, SICL requires feeding the spoken words and their labels separately to the encoder and decoder. 
SICL consistently reduces the word error rate (WER) irrespective of the Whisper model size or the specific dialect to adapt. 
The idea of $k$-nearest neighbours ($k$NN) can also be applied to in-context example selection \cite{C7}, which further improves SICL. 

In the rest of this paper: Sec.~\ref{sec:2} reviews the related work. Sec.~\ref{sec:3} presents the proposed SICL method. The experimental setup and results are given in Secs.~\ref{sec:4} and \ref{sec:5}. We conclude in Sec.~\ref{sec:6}.

\section{Related Work}
\label{sec:2}
\begin{figure*}[t]

\centering
\begin{minipage}[b]{0.425\linewidth}
  \centering  \hspace{-0.75cm}\centerline{\includegraphics[width=0.76\linewidth]{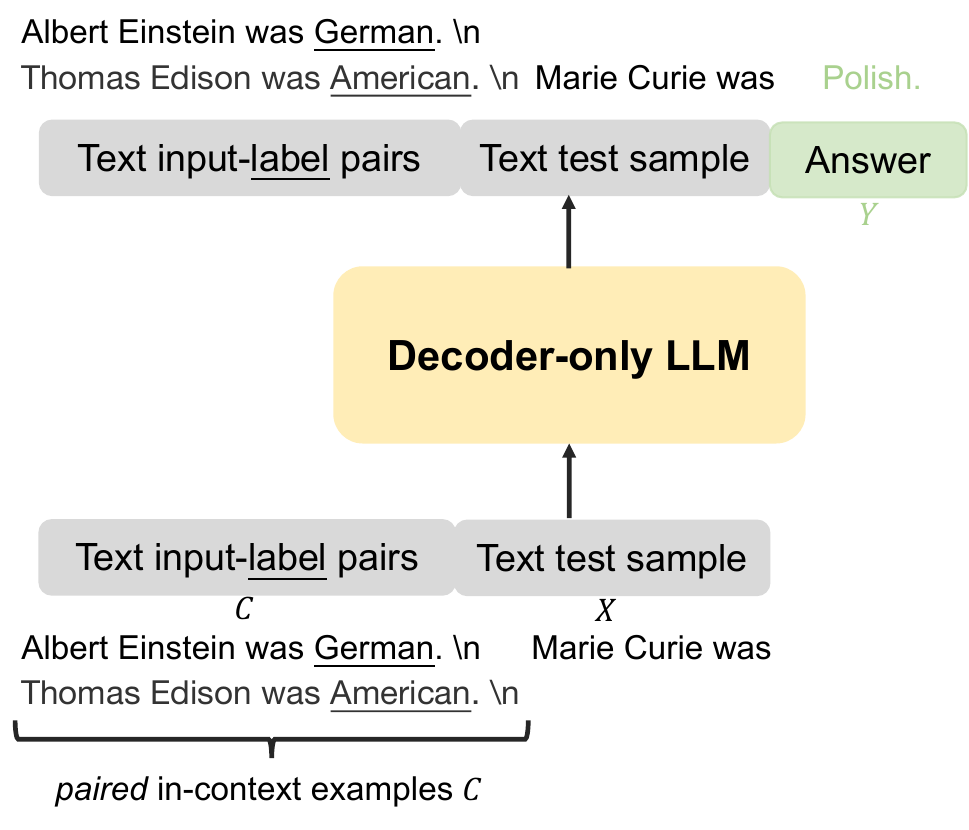}}
  \centerline{(a) Text-based ICL for LLMs.}\medskip
\end{minipage}
\begin{minipage}[b]{0.425\linewidth}
  \centering  \hspace{-0.15cm}\centerline{\includegraphics[width=1.23\linewidth]{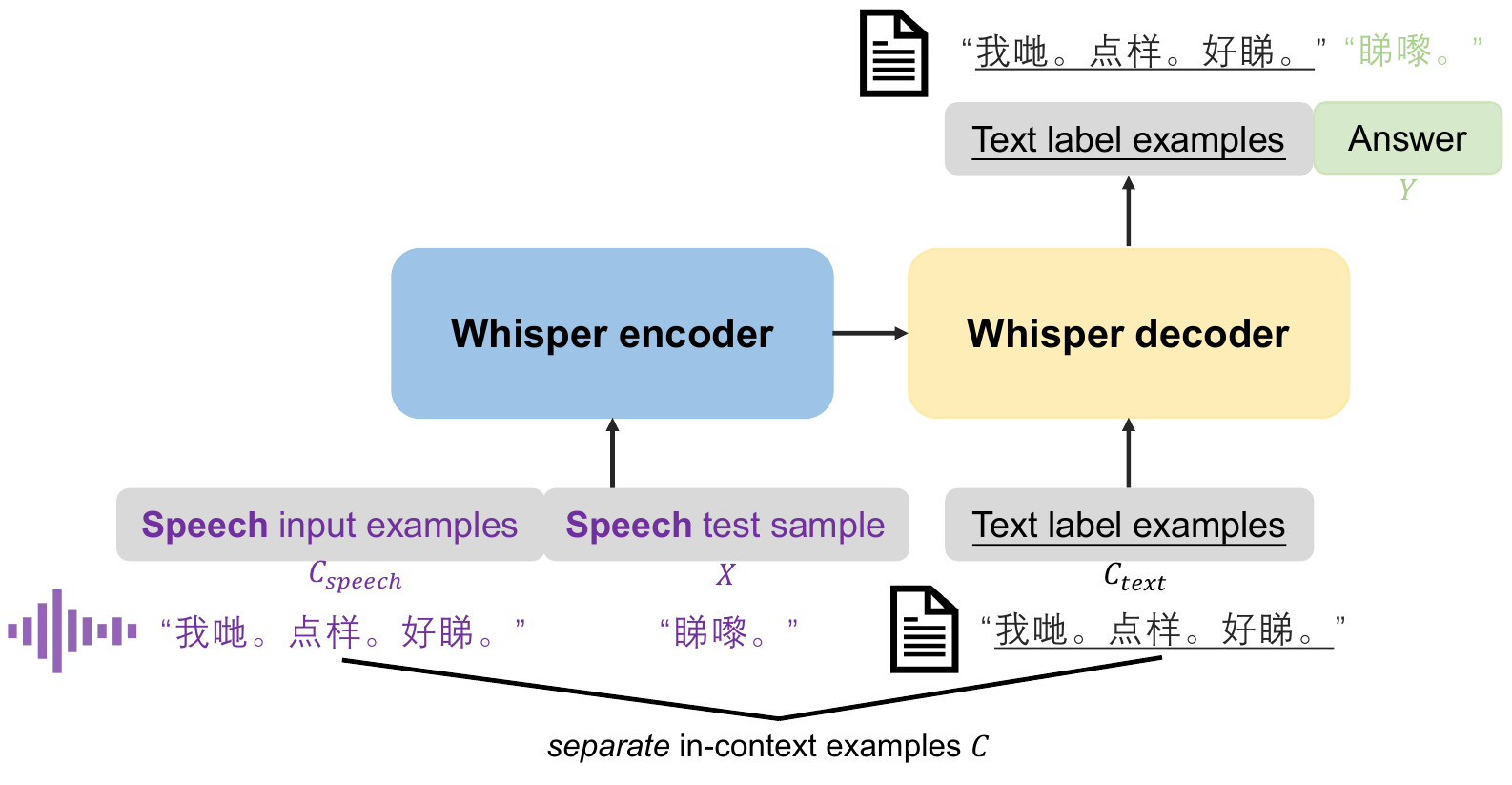}}
  \centerline{(b) SICL for end-to-end ASR (using Whisper as an example).}\medskip
\end{minipage}
\caption{Sketch maps of ICL for LLMs and SICL for end-to-end ASR. For ICL, text examples are fed to decoder-only LLM as the context, while for SICL, speech inputs and text labels of the in-context examples are fed separately and correspondingly into the encoder and decoder.}
\label{fig1}
\end{figure*}

\subsection{LLM and ICL studies for ASR}
How to apply LLM to ASR has been drawing increasingly more attention. Some studies used LLMs to rescore ASR output hypotheses \cite{C21,C22,C32}, which often requires finetuning LLMs. Min \textit{et al.} \cite{C23} tried to use LLMs' ICL ability to correct ASR errors. 
He \textit{et al.} \cite{C24} applied LLMs to spoken language understanding based on ICL. While \cite{C23,C24} use text-based LLM, our work studies  ICL in audio modality using Whisper as an example of large speech models. Peng \textit{et al.} \cite{C25} analysed Whisper's zero-shot generalisation by text-based prompting engineering. 
Efforts have been made to integrate LLMs with speech encoders into end-to-end ASR models \cite{yassir,wu2023}, and it's possible the resulting models have SICL abilities. 

\subsection{ASR for dialect speech applications}
Although different dialects may be categorised into the same language, they usually significantly differ in phonology and lexicon, which are similar to multilingual ASR with the same output token set. 
Each dialect can be modelled independently using a separate model \cite{C9,C10}, or jointly using a multi-dialect model \cite{C11,C12}. In multi-dialect ASR, dialect identifier can be detected \cite{C14} and used as part of the input \cite{C11,C13}. And exemplar-based speech enhancement method \cite{fin} is widely used for dialect adaption.
Multi-dialect ASR models also suffer from dialect bias \cite{C15,C16} due to an unbalanced training dataset, which can be handled by data augmentation \cite{C17} and balancing training corpora \cite{C18,C19}. 
When acquiring a new dialect, finetuning an existing ASR model is an effective approach \cite{C20}. In comparison, we use SICL to adapt Whisper without retraining the model.\\

\vspace{-0.3cm}
\section{Methodology}
\label{sec:3}
\subsection{The Whisper model}
\label{ssec:31}
Our work is based on Whisper \cite{C5}, a recently proposed series of large-scale speech models by OpenAI. 
The Whisper model family uses the encoder-decoder Transformer architecture \cite{C26} ranging from 39M parameters (tiny) to 1.55B parameters (large), and from an English-only version to a multilingual version covering 98 languages. Exquisite data formats were designed to enable Whisper to multitask with flexibility and robustness. Here we introduce two special tokens {\it prefix} and {\it prompt} 
\footnote{More information is available at \url{https://github.com/openai/whisper/discussions/117\#discussioncomment-3727051}}. 
The {\it prefix} token accepts a partial transcription for the current input audio segment, which is designed for cross-segment long-form audio ASR. The {\it prompt} token contains the text that can inform Whisper of certain jargon or style and supports further prompt engineering.
In the SICL method proposed in this section, {\it prefix} is used to present the labels of the spoken word in-context examples, and {\it prompt} is also designed to boost SICL performance. The \texttt{<|notimestamps|>} token is used to decode with text-only tokens, since it is observed that the text-only transcription without time stamps includes more accurate words for dialect ASR.

\subsection{Speech-based In-Context Learning (SICL)}
\label{ssec:32}
ICL refers to the ability that models can make inferences following the paradigm of the presented in-context examples. Regarding LLMs with decoder-only structures, such as GPT-3 \cite{C1}, LLaMa \cite{C28} and Vicuna \cite{C29}, the text examples can be presented directly as the dialogue history of the decoder.
Compared to LLMs, although end-to-end ASR models (\textit{e.g. }Whisper) also output word sequences in text modality, their input speech sequences are in audio modality and are much longer, which often requires to use an extra encoder structure to first convert the input speech into audio embeddings to avoid further increasing in the complexity of the decoders. Therefore, in our proposed SICL approach for end-to-end ASR, paired speech inputs and text labels are needed as the in-context examples, and the inputs and labels are required to be fed into the encoder and decoder separately and correspondingly. The pipelines of text-based ICL for LLMs and SICL for end-to-end ASR are visualised in Fig. \ref{fig1}.


Specifically, given a speech test sample $\mathbf{X}$, an ASR model $\Lambda$ decodes the most possible text sequence by $\hat{\mathbf{Y}}=\arg\max P(\mathbf{Y}|\mathbf{X},\Lambda)$. Let $\mathcal{C}=\{\mathcal{C}_\text{speech},\mathcal{C}_\text{text}\}$ be the in-context examples, 
where $\mathcal{C}_\text{speech}$ and $\mathcal{C}_\text{text}$ are the concatenations of a number of speech inputs and their paired text labels. 
With SICL, the probability of $\Lambda$ generating $\mathbf{Y}$ based on $\mathbf{X}$ and $\mathcal{C}$ can be rewritten as 
\begin{equation}
P_{\text{SICL}}(\mathbf{Y}|\mathcal{C},\mathbf{X},\Lambda) =P(\mathbf{Y}|\mathbf{E}_{\Lambda},\mathcal{C}_{\text{text}},\Lambda),
\end{equation}
where $\mathbf{E}_{\Lambda}$ is the audio embedding sequence extracted by the ASR encoder from the concatenation of $\mathcal{C}_{\text{speech}}$ and speech test sample $\mathbf{X}$, \textit{i.e.}
\begin{equation}
\mathbf{E}_{\Lambda}=\text{Encoder}_{\Lambda}(\text{concat}(\mathcal{C}_{\text{speech}},\mathbf{X})).
\end{equation}
In this paper, $\text{Encoder}_\Lambda(\cdot)$ is the encoder of Whisper model $\Lambda$, $\mathcal{C}_\text{text}$ are fed into the Whisper decoder via {\it prefix}. For comparison, conventional ICL for text LLMs can be presented as
\begin{equation}
P_{\text{ICL}}(\mathbf{Y}|\mathcal{C},\mathbf{X},\Lambda) =P(\mathbf{Y}|\mathcal{C},\mathbf{X},\Lambda),
\end{equation}
where $\mathbf{X}$ is the text test sample.  



For dialect ASR, $\mathcal{C}$ are the in-context examples consisting of speech from that dialect, which makes Whisper aware of the phonological deviation and specific dialectal words of the dialect. 

\subsection{$k$NN in-context example selection}
\label{ssec:33}
In SICL, how to select the in-context examples for test input $\mathbf{X}$ from a datastore $\mathcal{D}$ plays a critical role, which can be formulated as
\begin{equation*}
(\mathbf{X}_1,\mathbf{Y}_1),(\mathbf{X}_2,\mathbf{Y}_2),\ldots,(\mathbf{X}_k,\mathbf{Y}_k)=\text{SelectProcedure}(\mathbf{X},k,\mathcal{D}).
\end{equation*}
In this paper, the $k$NN method \cite{C7} is used as the selection procedure for SICL. First, generate the mean audio embedding $\mathbf{e}$ of the test speech $\mathbf{X}$, by averaging the audio embedding sequence $\mathbf{E}_{\Theta}=\text{Encoder}_{\Theta}(\mathbf{X})$ produced by the encoder of the example retrieval model $\Theta$ over time. 
Then $\mathbf{X}_1,\mathbf{X}_2,\ldots,\mathbf{X}_k$ along with their paired text labels are retrieved as the in-context examples of $\mathbf{X}$, whose mean audio embeddings $\mathbf{e}_1,\mathbf{e}_2,\ldots,\mathbf{e}_k$ have the smallest \textit{Euclidean distance} to $\mathbf{e}$. 
Note that the model $\Theta$ for in-context example retrieval can be different from the ASR model $\Lambda$.  
As a result,  
$\mathcal{C}_\text{speech}=\{\mathbf{X}_1,\mathbf{X}_2,\ldots,\mathbf{X}_k\}$ and $\mathcal{C}_\text{text}=\{\mathbf{Y}_1,\mathbf{Y}_2,\ldots,\mathbf{Y}_k\}$ consist of $k$ speech inputs and their paired text labels respectively.


\section{Experimental Setup}
\label{sec:4}
\subsection{Chinese dialects}
\label{sec:41}
Chinese speech has an extensive range of varieties.
In this work, two types of Chinese dialects, Chongqing and Guangzhou dialects, are studied. Chongqing dialect is a branch of Southwestern Mandarin, while Guangzhou dialect is a branch of Cantonese and is more different from Mandarin. Since Whisper was trained in standard Mandarin and Cantonese, it's intuitive to adapt Whisper, which already has some Chinese dialect knowledge , to a new Chinese dialect group by presenting in-context examples.

\subsection{Data}
\label{sec:42}


The 863 regional accent speech corpus (RASC863) \cite{C35a} commonly used for Chinese accent and dialect ASR studies is used as the data for in-context examples and test speech in this paper. 
Since the spontaneous and read speech in RASC863 were recorded for accented Mandarin, we only used the dialectal words of Chongqing and Guangzhou dialects from that dataset, which includes 2,994 and 1,990 words for Chongqing and Guangzhou dialects respectively. There are 200 speakers for each dialect region, balanced in terms of age, sex, and educational background.

To evaluate our approach to dialectal spontaneous speech, a private dialectal conversation dataset obtained from iFLYTEK is also used as a test set.  
The iFLYTEK dataset contains voice-changed telephone conversations in the Chongqing and Guangzhou dialects, both with around 85 hours of duration. For each dialect, we randomly chose 5,000 utterances as our spontaneous test set, with around 7 hours duration.

\subsection{Models, prefixes and prompts}
\label{sec:43}
\begin{CJK*}{UTF8}{gkai}
Whisper models with four different sizes are used, which include base (B), small (S), medium (M), and large (L) with $74$M, $244$M, $769$M, and $1.55$B parameters. The language identity (LID) input of Whisper is not set unless explicitly specified, and ``LID'' means to set the LID to ``zh'' (Chinese). 
The full stop ``。'' in written Chinese is used as the delimiter between each in-context example in our study. A {\it prompt} ``识别方言'' (recognize dialect speech) is used to boost the performance by informing Whisper that it's a dialect ASR task. This performance gain may be attributed to the similar {\it prompt} with which Whisper was trained on dialect data. The punctuations and duplicate tokens produced by Whisper were removed based on automatic rules from the transcriptions before scoring. The WERs are calculated using the SCTK toolkit. 

\end{CJK*}

\begin{table}[!h]
\caption{\%WERs on RASC863 dialectal word dataset. 
$k$ indicates the number of in-context examples used in SICL, and $k=0$ is the baseline without SICL and LID. The in-context examples are randomly selected three times from the relevant datastore, which includes the RASC863 dialectal words of the same dialect and same speaker. The average WERs are reported.}
\setlength{\tabcolsep}{4pt}
\centering
\begin{tabular}{cc|cccc|cccc}
\toprule
 & & \multicolumn{8}{c}{Dialect \& Whisper model size} \\ [0.1cm]
\multicolumn{2}{c|}{Setting} & \multicolumn{4}{c|}{Chongqing} & \multicolumn{4}{c}{Guangzhou} \\ [0.1cm]
& & B & S & M & L & B & S & M & L \\
\midrule
\multicolumn{2}{c|}{LID} & 75.3 & 83.4 & 69.8 & 68.7 & 85.6 & 92.9 & 65.8 & 66.2 \\ 
\midrule
\multicolumn{2}{c|}{$k=0$} & 90.3 & 89.5 & 78.4 & 84.7 & 118.9 & 107.3 & 85.3 & 89.3 \\ 
\multicolumn{2}{c|}{$k=1$} & 75.9 & 73.2 & 64.5 & 67.1 & 90.4 & 86.7 & 54.0 & 61.7 \\
\multicolumn{2}{c|}{$k=2$} & 72.3 & 66.9 & 57.7 & 56.1 & 73.2 & 71.1 & 37.0 & 38.3 \\
\multicolumn{2}{c|}{$k=3$} & 70.2 & 64.6 & 55.5 & 52.7 & 66.0 & 60.6 & 32.7 & 31.2 \\
\multicolumn{2}{c|}{$k=4$} & 68.7 & 62.9 & 53.6 & \textbf{51.0} & 63.1 & 56.2 & 30.3 & \textbf{28.8} \\
\bottomrule
\end{tabular}
\label{t1}
\end{table}

\begin{table*}[ht]
\caption{\%WERs on RASC863 dialectal word dataset using SICL with different in-context example selection and presentation settings. The example retrieval models $\Theta$ and inference models $\Lambda$ are both Whisper models of different sizes. The number of in-context examples is chosen to be $k=4$. The example datastore $\mathcal{D}$ is the RASC863 dialectal word dataset of the same dialect and same speaker. ``far to near'' and ``near to far'' mean whether the in context-examples are presented in the order of reducing the distance to the test speech, or \textit{vice versa}. }
\setlength{\tabcolsep}{4pt}
\centering
\begin{tabular}{cc|cccc|cccc}
\toprule
\multirow{3.5}{*}{\makecell{Example\\Selection\\Method}} & \multirow{3.5}{*}{\makecell{Example\\Present\\Order}} & \multicolumn{8}{c}{Dialect \& Whisper model size} \\ [0.1cm]
 &  &\multicolumn{4}{c|}{Chongqing} & \multicolumn{4}{c}{Guangzhou} \\ [0.1cm]
& & B & S & M & L & B & S & M & L \\
\midrule
Random & & 68.7±0.56 & 62.9±0.14 & 53.6±0.33 & 51.0±0.21 & 63.1±0.62 & 56.2±0.26 & 30.3±0.25 & 28.8±0.65\\
\midrule
\multirow{2}{*}{$k$NN, $\Theta=\text{B}$} & far to near & 67.5 & 62.4 & 52.3 & \textbf{49.2} & 58.5 & 52.7 & 26.7 & 25.0\\
& near to far & 67.8 & 62.1 & 52.4 & 49.6 & 60.1 & 51.2 & \textbf{26.4} & 24.8\\
 \multirow{2}{*}{$k$NN, $\Theta=\text{S}$} & far to near & 66.3 & \textbf{61.5} & 52.6 & \textbf{49.2} & 58.3 & 51.9 & 27.6 & \textbf{24.5}\\
 & near to far & 66.1 & 61.6 & 52.9 & 49.5 & 57.3 & 52.8 & 27.7 & 24.8\\
 \multirow{2}{*}{$k$NN, $\Theta=\text{M}$} & far to near & 66.9 & 63.2 & 53.9 & 50.6 & 58.2 & 53.6 & 29.0 & 26.7\\
 & near to far & 67.4 & 63.6 & 53.8 & 50.7 & 57.2 & 54.3 & 28.3 & 25.3\\
 \multirow{2}{*}{$k$NN, $\Theta=\text{L}$} & far to near & 65.9 & 61.6 & \textbf{52.2} & 49.7 & \textbf{55.3} & \textbf{49.0} & 27.5 & 24.8\\
 & near to far & \textbf{65.8} & 62.1 & 52.6 & 49.7 & 55.4 & 49.7 & 27.9 & 24.8\\
\bottomrule
\end{tabular}
\label{t2}
\end{table*}

\begin{table}[h]
\caption{\%WERs of spontaneous speech using the iFLYTEK test set. $k$ is the number of in-context examples, selected by $k$NN with the example retrieval model $\Theta=\text{L}$ .The example presentation order is "far to near". The example datastore is the RASC863 dialectal word dataset of corresponding dialect.}
\vspace{0.1cm}
\setlength{\tabcolsep}{4pt}
\centering
\begin{tabular}{c|cccc|cccc}
\toprule
 & \multicolumn{8}{c}{Corpus \& Whisper model size} \\ [0.05cm]
Setting & \multicolumn{4}{c}{iFLYTEK Chongqing} & \multicolumn{4}{c}{iFLYTEK Guangzhou}\\ [0.05cm]
& B & S & M & L & B & S & M & L \\
\midrule
LID & 75.9 & 68.1 & 61.2 & 56.9 & 60.0 & 50.8 & 42.4 & 48.3 \\ 
LoRA & 76.3 & 61.3 & 51.5 & 51.0 &  51.4 & 52.4 & 33.7 & 21.8\\ 
\midrule
$k=0$ & 80.2 & 68.5 & 63.3 & 60.5 & 66.0 & 53.7 & 43.5 & 52.1\\ 
$k=1$ & 77.1 & 67.5 & 59.7 & 57.5 & 59.3 & 51.4 & 39.5 & 39.9\\
$k=2$ & 75.2 & 64.5 & 57.1 & 55.3 & 54.9 & 49.5 & 36.7 & 34.4\\
$k=3$ & 74.0 & 63.6 & 55.8 & 54.3 & 54.4 & 47.8 & 34.8 & 32.8\\
$k=4$ & 73.2 & 62.9 & 54.8 & 53.3 & 53.0 & 47.6 & 34.0 & 30.7\\
\bottomrule
\end{tabular}
\label{t3}
\end{table}

\section{Experimental Results}
\label{sec:5}
\subsection{Speech-based in-context learning}
\label{ssec:51}
Table \ref{t1} shows the WERs on the RASC863 dialectal word dataset. 
By simply informing the LID, Whisper received an 17.9\% relative WER reduction averaging dialects and model sizes (comparing the LID row to $k=0$ row). This indicates that correct priori knowledge can benefit the inference of Whisper, as expected. The SICL method provides WER reductions in most cases when the number of in-context examples $k$ is greater than or equal to 2. 
With $k=4$, SICL can outperform decoding with LID zh by 17.5\%, 32.0\%, 38.6\%, and 41.1\% relative WER reductions on Whisper model sized B, S, M, and L respectively. 
Our results are consistent with previous findings with LLMs \cite{C27} that the accuracy of ICL improves when the number of in-context examples or model size increases. It is notable that since Whisper model B can already gain a 26.3\% relative WER reduction on the Guangzhou dialect with SICL, SICL does not necessarily require a very large model size to be effective.


\subsection{$k$NN-based example selection}
\label{ssec:52}
Table \ref{t2} shows the WERs of Whisper models on the RASC863 dialectal word dataset with different settings of SICL.
Regarding the order of presentation, in-context examples can be presented in the order of reducing the distance to the test speech, referred to as ``far to near'', or \textit{vice versa}, referred to as ``near to far''. 
The $k$NN in-context example selection method outperforms random selection in most cases, apart from when using Whisper model M as the example retrieval model ($\Theta=\text{M}$), leading to 6.4\% relative WER reduction on average when using $\Theta=\text{L}$ and the ``far to near'' order. 
This reveals that the examples closer to the test speech contain more useful contextual information for dialect adaptation.

The use of different example retrieval models $\Theta$ is also investigated. 
The results suggest that serving as a good example retrieval model ($\Theta$) and being adapted to generate better ASR transcriptions ($\Lambda$) are independent abilities, and similar findings have been observed on LLMs \cite{C30}. 
Compared to the example selection method, the influence of presentation order is secondary, meaning the choice of examples is more important than how they are presented.  Our finding is validated on the iFLYTEK spontaneous speech test
set, and the WERs are reported in Table \ref{t3}. The low-rank adaptation (LoRA) method \cite{C33hu2021lora} is compared as a strong adaptation baseline (row 2), in which the attention layers of Whisper are adapted using LoRA with rank of 4 and scale of 16 based on the RASC863 dialectal words data. For this setting, 
the datastore $\mathcal{D}$ is chosen to be the  RASC863 dataset of the nearest speaker in corresponding dialect. The nearest speaker is determined by the \textit{Euclidean distance} between the average speaker audio embedding and test speech. Then $k$NN method is used to choose the in-context examples from that speaker. Compared to decoding with LID zh (row 1), the SICL method achieves on average a relative 12.8\% WER reduction. With few words presented, SICL can achieve comparable WERs to the use of LoRA, without requiring any gradient descent or model parameter update. 
Since the datastore used in Table \ref{t3} is still the dialectal words in RASC863, which is out-of-domain in terms of both the speech type and the channel of recordings when compared to the iFLYTEK spontaneous speech test set. This reveals that SICL learned the phonological variances and lexical nuances in the dialect and is robust against the changes in speaker and channel.

\subsection{Levels of adaptation}
\label{ssec:53}
Finally, the effectiveness of SICL is investigated at different levels of adaptation according to how datastores with the RASC863 dialectal words for the in-context example selection are constructed.  
Table \ref{t4} shows the results of the ablation studies with the $k$NN example selection method.  {\it same speaker same dialect} is the default datastore same as the experiment of Table \ref{t2}.
For {\it different speaker same dialect}, the datastore consists of the dialectal words from the same dialect but different speaker, which is controlled to achieve language-level adaptation. This specific speaker is the nearest speaker measured by the {\it Euclidean distance} between the average speaker audio embedding and test speech.
The same words are avoided to be selected to ensure fair comparisons. 
For 
{\it same speaker different dialect}, the datastore consists of the read speech of standard Mandarin with a slight accent from the same speaker, which is possible since the RASC863 corpus includes such read speech. 
Note the in-context examples are read speech of short sentences rather than isolated dialectal words.  

\begin{table}[!h]
\caption{\%WERs of ablation experiments on RASC863 dialectal word dataset with different settings for datastore. Inference models $\Lambda$ are Whisper models of different sizes. The example retrieval model is $\Theta=\text{L}$ and the example presentation order is the "far to near". The number of in-context examples is $k=4$. }

\vspace{0.05cm}
\setlength{\tabcolsep}{4pt}
\centering

(a) Chongqing dialect
\vspace{0.05cm}

\begin{tabular}{c|cccc}
\toprule
\multirow{2}{*}{Datastore ($\mathcal{D}$)} & \multicolumn{4}{c}{Whisper model size} \\ [0.05cm]
 & B & S & M & L\\
\midrule
same speaker same dialect & 65.9 & 61.6 & 52.2 & 49.7   \\ 
\midrule
different speaker same dialect & 68.9 & 63.4 & 55.7 & 52.9 \\
same speaker different dialect& 68.4 & 66.0 & 66.2 & 61.1 \\
\bottomrule
\end{tabular}

\vspace{0.1cm}

(b) Guangzhou dialect
\vspace{0.05cm}

\begin{tabular}{c|cccc}
\toprule
\multirow{2}{*}{Datastore ($\mathcal{D}$)} & \multicolumn{4}{c}{Whisper model size} \\ [0.05cm]
 & B & S & M & L\\
\midrule
same speaker same dialect & 55.3 & 49.0 & 27.5 & 27.3  \\ 
\midrule
different speaker same dialect & 61.9 & 57.7 & 33.3 & 30.5 \\
same speaker different dialect& 92.2 & 90.4 & 80.9 & 81.3 \\
\bottomrule
\end{tabular}
\label{t4}
\end{table}
Our findings emphasise the primacy of language-level adaptation. It's pertinent to note that the Chongqing dialect belongs to the Mandarin dialect group, whereas the Guangzhou dialect markedly deviates from Mandarin. Leveraging Mandarin in-context examples enhances recognition for Chongqing dialect speech, but paradoxically diminishes performance for the Guangzhou dialect, falling even below the no-context baseline (row 1 in Table~\ref{t1}). This suggests that while proximate dialects can be useful in context exemplars, incongruous examples might detrimentally affect Whisper's inference. Interestingly, the performance degradation is more pronounced in Whisper models M and L, implying that larger models, while predisposing to capitalise on contextual cues for ASR transcription, are more susceptible to deviations introduced by extraneous context.

\section{CONCLUSION}
\label{sec:6}
In this work, the SICL method is proposed for ASR test-time adaptation without any gradient descent based on the Whisper models, and a $k$NN-based example selection is used to enhance SICL. Experimental results on Chinese dialect ASR and adaptation show that SICL consistently and considerably outperforms the decoding with correct LID baseline with an average relative 36.4\% WER reduction, underscoring Whisper's intrinsic ICL abilities. Even with an out-of-domain datastore, SICL can robustly improve ASR performance.
Further analysis reveals that language-level adaptation is more effective than speaker-level for SICL. Our future work includes exploring the mechanism of SICL and studying SICL in more speech tasks, with longer contexts and other large speech models. 

\clearpage
\small
\bibliographystyle{IEEEbib}
\bibliography{refs}

\end{document}